\documentstyle[12pt,a4]{article}
\newcommand{\be}{\begin{equation}}
\newcommand{\ee}{\end{equation}}
\begin{document}
\vspace{20.0mm} 
\title{Magnetic field inversion in vortices in multilayers}
   
\author{{\Large Stavros Theodorakis and Epameinondas Leontidis}\\
         \\
            Dept. of Natural Sciences, University of Cyprus\\
         {\it P.O. Box 537, Nicosia 1678, Cyprus}}

\maketitle

\vspace{35.0mm}
\begin{abstract}

\noindent
We present a description of very dense vortex lattices in highly
anisotropic multilayers, for high fields parallel to the layers.
We show that a magnetic field inversion can occur away from the
center of a vortex, provided the layers are sufficiently far
apart.

\end{abstract}

\newpage
The structure of vortices in layered superconductors presents a
fascinating array of peculiar characteristics, depending on the
direction of the external magnetic field. Some of these have been
studied quite some time ago, while others more 
recently~\cite{Car91,Theo95}. The case where the external field
is
parallel to the layers is an especially interesting one. It was
shown numerically~\cite{Koy91}, for example, that in layered
superconductors with inequivalent layers immersed in a magnetic
field parallel to the layers, the field around a vortex can be
inverted. The inversion can be achieved only when the in-plane
penetration depth of the weakly superconducting layers is much
greater than that of the strongly superconducting layers. This
work was based on the assumption of an essentially uniform order
parameter amplitude along the layers. 

\vspace*{3mm}
The present paper aims to show that the above mentioned field
inversion for vortices parallel to the layers can be a really
generic effect in highly anisotropic multilayers, where the
superconducting structure undulates along the z-axis, as long as
the period of undulation is large enough, and as long as the
applied magnetic field is high enough. We show in particular
that if the period of undulation d of the superconducting
structure is much larger than a penetration
depth $\lambda$ along the z-axis, then the field of a vortex is
inverted away from the center of the vortex,
for sufficiently high external fields. 

\vspace*{3mm}

What happens essentially is that the field $h$ of each vortex can
go quite fast along the z-axis from a nonzero value $h(0)$ at the
center of the
vortex to the zero value of the Meissner state, provided the
penetration depth $\lambda$ is small enough. In fact, for
sufficiently large
values of $d/\lambda$ the field $h$ could go to zero within
a distance much shorter than d, where d is the distance between
layers. Indeed, it will do so for dense vortex lattices with
d$\gg\lambda$, where
there are vortex centers halfway between the layers, because the
superconducting layers expel the magnetic field, limiting it to
the interlayer regions. It turns out now
that the value $h(0)$ of the field at the center of the vortex,
which lies halfway between layers, 
is
approximately equal to the value H of the
external field, for large H. Ideally, it
would be zero on the superconducting layer due to screening, if
the field can vary rapidly enough. This rapid variation is
ensured if
d is much greater than $\lambda$, unlike the case of the usual
superconductors or of high temperature superconductors, where the
field changes very slowly on the scale of d. Thus, if $d/\lambda$
is large enough, the field will start out at a high value halfway
between layers and drop abruptly within just a few $\lambda$'s.
It will have therefore a large negative initial slope, which
makes it shoot down very abruptly. It will thus overshoot past
zero towards negative values, before the existence of
superconductivity on the layer forces it to come back up towards
zero. The field therefore will decrease very rapidly from a large
$h(0)$ to a
negative value, then go to zero within the superconducting
layer, and then rise again in the next interlayer region, where
the external field can penetrate easily. A field inversion can
occur therefore, if
$d/\lambda$ and H are sufficiently large, because the large
negative slope of $h$ makes it drop down to negative values
just before
becoming zero around the superconducting layer.

\vspace*{3mm}
In order to demonstrate the above mentioned field inversion, we
shall use a very general model that takes into account the
existence of a nonzero order parameter between the
layers~\cite{Theo93}. We shall assume in particular that the
Gibbs free energy of the multilayer in an external field H
parallel to the x-axis is
\begin{eqnarray}
\int\int\int\,dx\,dy\,dz\Bigl[a(z)|\Psi|^{2}
+\beta|\Psi|^{4}/2+\frac{\hbar^{2}}{2m}\bigg|-i
\nabla_{\parallel}\Psi-\frac{2e}{\hbar
c}A_{\parallel}\Psi\bigg|^{2}\nonumber\\
+\frac{\hbar^{2}}{2M}\bigg|-i\frac{\partial\Psi}{\partial
z}-\frac{2e}{\hbar
c}A_{z}\Psi\bigg|^{2}
+\frac{1}{8\pi}(\nabla\times A-
H)^{2}\Bigr].
\end{eqnarray}
Here z is the direction normal to the layers. 

\vspace*{3mm}
We shall measure x, y, z in terms of d, the period of the
superconducting structure along the z-axis. We render $a(z)$
dimensionless by taking out a dimensionful constant $\alpha$, so
that $a(z)/\alpha$$=\alpha(z)$. We measure $\Psi^{2}$ in units
of $\alpha/\beta$, the vector potential A in units of $\hbar
c/2ed$, the magnetic fields in units of $\hbar c/2ed^{2}$, and
the energies in units of $d^{3}\alpha^{2}/\beta$. Thus the Gibbs
free energy density takes the dimensionless form
\begin{eqnarray}
g=\Bigl[\alpha(z)|\Psi|^{2}
+|\Psi|^{4}/2+\nu\Gamma^{2}|-i\nabla_{\parallel}\Psi-
A_{\parallel}\Psi|^{2}+\nonumber\\
\nu|-i\frac{\partial\Psi}{\partial z}-
A_{z}\Psi|^{2}
+\nu\Gamma^{2}\frac{\lambda^{2}}{d^{2}}(\nabla\times A-
H)^{2}\Bigr], 
\end{eqnarray}
where $\nu=\hbar^{2}/2M\alpha d^{2}$, $\Gamma^{2}=M/m$ and
$\lambda^{2}=mc^{2}\beta/16\pi e^{2}\alpha$. The penetration
depth $\lambda$ should not be confused with the experimentally
measured penetration depth. Note furthermore that $\lambda$
depends on the choice of the dimensionful constant $\alpha$,
which is unspecified so far, while $\lambda^{2}/\nu d^{2}$ does
not. 

\vspace*{3mm}
In this paper we shall examine very dense vortex lattices. In
such lattices the vortex cells are practically rectangular.
Vortices
exist in every interlayer spacing, but in a squeezed triangular
formation, due to vortex repulsion (see Fig. 1). The vortices in
the vortex lattice are parallel to the x-axis, because the
external
field is assumed to be along that axis. The order
parameter $\Psi$ becomes maximum on each layer, the distance
between successive layers being d, but vortices will appear only
every two layers along the z-axis, due to their mutual repulsion.
Thus each rectangular vortex cell will have a length of 2L along
the y-axis, and 2d along the z-axis. As the field H increases,
the vortex cells get more narrow, in other words L decreases, but
the structure along the z-axis remains unchanged. Indeed, due to
their mutual repulsion the distance between two vortices along
the z-axis cannot be reduced below 2d. We have thus a triangular
lattice, with vortices situated in every interlayer spacing. The
spatial periodicity of the vortex lattice results in the
quantization of the magnetic flux contained in each unit cell.
We anticipate therefore that $L\propto 1/H$.

\vspace*{3mm}
We expect each vortex to be axisymmetric very close to its
center.
The vector potential and the magnetic field vary typically over
distances of the order of $\lambda$ along the z-axis, but of
order $\Gamma\lambda$ along the y-axis. We shall work
consistently in the limit $d\gg\lambda$, and
$\Gamma\rightarrow\infty$. In other words, our results will hold
for highly anisotropic multilayers, with the layers sufficiently
far apart. 

\vspace*{3mm}
Very close to the vortex center, the field is axisymmetric. Thus,
in the vicinity of the center, we expect the lines of constant
field to be ellipses centered at the vortex center, i.e.
$h=h(\sqrt{z^{2}+y^{2}/\Gamma^{2}})$. Since the magnetic field
decays to zero pretty fast, within just a few $\lambda$'s along
the z-axis, the various vortices do not overlap in the z
direction. Note also that
$\alpha(z)\approx\alpha(\sqrt{z^{2}+y^{2}/\Gamma^{2}})$ for
$|z|\gg L/\Gamma > |y|/\Gamma$. Thus, if $L\ll\Gamma$,
$\alpha(z)$ is axisymmetric practically everywhere. It is thus
reasonable to assume that the axisymmetry
of the magnetic field will be approximately true over most of the
cell, as long as $\Gamma$ is very large. We can easily verify
that the ansatz
\be
A_{y}=-\frac{\sin\phi}{\Gamma}A(\rho),
\ee
\be
A_{z}=\cos\phi A(\rho),
\ee
with $A_{x}=0$, can lead to such an axisymmetric single vortex
field. The
anisotropic polar coordinates $\rho$, $\phi$
on the y-z plane are defined through the equations
\be
\rho=\sqrt{z^{2}+y^{2}/\Gamma^{2}},
\ee
\be
\sin\phi=\frac{z}{\rho},
\ee
\be
\cos\phi=\frac{y}{\Gamma\rho},
\ee
assuming that the origin is the center of the vortex.

\vspace*{3mm}
We shall therefore assume that this ansatz for the vector
potential holds over the whole cell. Since $\alpha(z)$ is really
$\alpha(\rho)$ for very anisotropic multilayers
($\Gamma\rightarrow\infty$), provided $|y|\ll\Gamma$, the order
parameter amplitude and the magnetic field are essentially
functions of
$\rho$ only, justifying thus the ansatz of Eqs. (3) and (4). 

\vspace*{3mm}
The magnetic field of the vortex will be along the x-axis:
\be
h(\rho)=\frac{1}{\Gamma\rho}\frac{\partial Q}{\partial\rho},
\ee
with 
\be
Q(\rho)=-1+\rho A(\rho).
\ee
The approximate axisymmetry of the magnetic field and of
$\alpha(z)$ imply that the magnitude of the order parameter is
approximately axisymmetric as well: 
\be 
\Psi(\rho,\phi)=\psi(\rho)e^{i\phi}.
\ee

\vspace*{3mm}
The Gibbs free energy density takes the following form, in terms
of $\psi$ and Q:
\be
g=\alpha(z)\psi^{2}
+\frac{1}{2}\psi^{4}+\nu(\frac{\partial\psi}
{\partial\rho})^{2}+\frac{\nu}{\rho^{2}}Q^{2}\psi^{2}
+\frac{\nu\lambda^{2}}{d^{2}}\bigl(\frac{1}{\rho}
\frac{\partial Q}{\partial\rho}-\Gamma H\bigr)^{2}.
\ee

\vspace*{3mm}
The superconducting structure is represented by $\alpha(z)$. If
this structure has a period d, then $\alpha(z)$ is periodic with
a period d. We expect that the vortices are situated halfway
between the layers. Therefore $\alpha(z)$ will have an extremum
at the center of the vortex. Combined with the periodicity, this
implies that $\alpha(z)$ is an even function of z. 

\vspace*{3mm}
We repeat that since $\Gamma\rightarrow\infty$, we can assume
that $\alpha(z)$ is really $\alpha(\rho)$, as long as
$|y|<L\ll\Gamma$. Thus $\psi$, $Q$, $A$ and $h$ are essentially
functions of $\rho$. 

\vspace*{3mm}
For a very dense vortex lattice, i.e. a high external field, the
unit cell will be a truncated ellipse, with $y\ll\Gamma$ within
it (see Fig. 2), and it will resemble a rectangle with sides $2L$
and 2, where
$L\ll\Gamma$. In that case
$|z|\leq\rho\leq$$\sqrt{z^{2}+L^{2}/\Gamma^{2}}$$\approx|z|$,
hence $\rho\approx|z|$. The flux in the unit cell will then be
$2L\cdot
2\int_{0}^{1}h\,d\rho$$=2\pi$. The field
varies very little along the y-axis when the y coordinate is much
less than $\Gamma$, so when $L$ has become quite small the
field $h$ will be varying along the z-direction only, remaining
almost constant along the y-direction. The order parameter will
also be a function of z only, remaining practically constant in
the y-direction. We can say equivalently that $\psi$ is a
function of $\rho$ only.

\vspace*{3mm}
The Gibbs free energy will be the integral of $g$ over the area
of the truncated cell of Fig. 2. Thus the Gibbs free energy per
unit x-length will equal
\be
\int_{0}^{L/\Gamma}d\rho\,\,2\pi\Gamma\rho g
+\int_{L/\Gamma}^{1}d\rho\,\,\Gamma\rho g\bigl[2\pi
-4\cos^{-1}\bigl(\frac{L}{\Gamma\rho}\bigr)\bigr].
\ee
The equations that minimize this functional for $0<\rho<L/\Gamma$
are:
\be
\alpha(z)\psi+\psi^{3}+\frac{\nu
Q^{2}\psi}{\rho^{2}}=\nu\frac{\partial^{2}\psi}{\partial\rho^{
2}}+\frac{\nu}{\rho}\frac{\partial\psi}{\partial\rho},
\ee
\be
\frac{\partial}{\partial\rho}\bigl(\frac{1}{\rho}\frac{\partial
Q}{\partial\rho}\bigr)=\frac{d^{2}}{\lambda^{2}}
\frac{Q\psi^{2}}{\rho}
\ee
Indeed, these
equations minimize the first integral in Eq. (12). Clearly, since
$L\ll\Gamma$, these equations really hold only at the vicinity
of the vortex center. We can easily show then that $Q(0)=-1$ and
$\psi(0)=0$. Furthermore,
$\partial h/\partial\rho$ is zero at $\rho=0$ and $\psi$ is
linear in $\rho$ near the origin.

\vspace*{3mm}
Let us now look at the case $L/\Gamma<\rho<1$. If $L\ll\Gamma$,
then the second integral of Eq. (23) can be approximated by the
integral $\int_{L/\Gamma}^{1}4Lg\,d\rho$, because $\cos^{-
1}(L/\Gamma\rho)$$\approx(\pi/2)-(L/\Gamma\rho)$ on most of the
interval $[L/\Gamma,\,1]$. Thus we have to find the field
equations that minimize $\int_{L/\Gamma}^{1}g\,d\rho$. These
equations are
\be
\alpha(\rho)\psi+\psi^{3}+\frac{\nu Q^{2}\psi}{\rho^{2}}=
\nu\frac{\partial^{2}\psi}{\partial\rho^{2}},
\ee
where $\rho\approx|z|$ within the cell, and
\be
\frac{\partial}{\partial\rho}\Bigl(\frac{\Gamma h-\Gamma
H}{\rho}\Bigr)=\frac{d^{2}}{\lambda^{2}}\frac{Q\psi^{2}}
{\rho^{2}}.
\ee
This last equation can be rewritten in the form
\be
\rho\Gamma\frac{\partial h}{\partial\rho}-\Gamma h+\Gamma
H=\frac{d^{2}}{\lambda^{2}}Q\psi^{2}.
\ee
We stress again that Eqs. (15) and (17) hold $only$ in the
interval $L/\Gamma<\rho<1$.

\vspace*{3mm}
Let us now discuss the boundary conditions. We already said that
$Q(0)=-1$ at the center of the vortex. Also $\psi(0)=0$. Since
the vortex is situated halfway between the layers, and since
$\psi$ varies little along the y-axis, due to the small $L$ and
the large $\Gamma$, the
order parameter must be zero on all the planes that are located
halfway between neighboring superconducting layers. But the edges
of the cells lie also on such planes. Hence $\psi(1)=0$. The
field $h$ has to fit with that of the next cell, so we must also
have that $\partial h/\partial\rho$ be zero at the
edge of the cell, i.e. $\dot h(1)=0$. Combined with the boundary
condition $\psi(1)=0$ and with Eq. (17) this implies that
$h(1)=$H. The boundary conditions for Eqs. (15) and (17) are thus
$\psi(1)=0$, $h(1)=$H at the edge of the cell. 

\vspace*{3mm}
Since the interval [0, $L/\Gamma$] is quite small, we can
consider the boundary conditions $\psi(0)=0$, $Q(0)=-1$ as
appropriate for Eqs. (15) and (17) as well, even though those
equations
apply strictly to the interval [$L/\Gamma$, 1] $only$.

\vspace*{3mm}
We conclude then that Eqs. (15) and (17) describe fully the dense
vortex lattice if the y-length of the cell is much smaller than
$\Gamma$, subject to the boundary conditions $\psi(0)=\psi(1)=0$,
$Q(0)=-1$, $h(1)=$H. 

\vspace*{3mm}
We shall explore the qualitative consequences of this description
through a simple variational model, before conducting a more
careful numerical study. Thus, we shall adopt the trial order
parameter
\be
\psi^{2}(\rho)=\psi_{0}^{2}\delta(\rho-\frac{1}{2}),
\ee
where $\psi_{0}$ is determined by the details of $\alpha(z)$, and
we shall solve for $Q$.

\vspace*{3mm}
Equation (16), along with the boundary condition $Q(0)=-1$,
yields
\be
Q(\rho)=-1+\Gamma H\rho^{2}/2+\Gamma\kappa\rho^{3}/3
\ee
for $0<\rho<1/2$, as well as
\be
h(\rho)=H+\mu\rho
\ee
for $1/2<\rho<1$, $\kappa$ and $\mu$ being integration constants.

\vspace*{3mm}
The boundary condition $h(1)=$H, i.e. $\dot h(1)=0$, gives
$\mu=0$. Thus we get
\begin{eqnarray}
h(\rho)=H+\kappa\rho,\,\,\,\,\,if\,\,\,0<\rho<1/2\nonumber\\
=H,\,\,\,\,\,if\,\,\,1/2<\rho<1.
\end{eqnarray}
Let us now integrate Eq. (16) from $\frac{1}{2}-\epsilon$ to
$\frac{1}{2}+\epsilon$, $\epsilon$ being a positive number
tending to zero. Using the ansatz of Eq. (18) we get the boundary
condition at $\rho=1/2$:
\be
\Gamma h(\frac{1}{2}+)-\Gamma h(\frac{1}{2}-)=2\omega
Q(\frac{1}{2}),
\ee
with $\omega=d^{2}\psi_{0}^{2}/\lambda^{2}$. We combine now Eqs.
(19), (21) and (22) to get 
\be
\Gamma\kappa=\frac{3\omega(8-\Gamma H)}{6+\omega}.
\ee
Hence 
\be
\Gamma h(\frac{1}{2}-)=\Gamma H+\frac{\Gamma\kappa}{2}=
\frac{\Gamma H(12-\omega)+24\omega}{12+2\omega},
\ee
while $h(\frac{1}{2}+)=$H. Hence $h$ changes discontinuously
across the superconducting layer. A similar change was presented
in [2].

\vspace*{3mm}
We now note that $h(\frac{1}{2}-)$ will be negative if and only
if $\omega>12$ and $\Gamma H>24\omega/(\omega -12)$. Hence this
naive model shows that the field $h$ of the vortex will be
inverted if the external field H is sufficiently high, and if
$d/\lambda$ is sufficiently large.

\vspace*{3mm}
We shall now obtain further results with the aid of numerical
work done with the choice
\be
\alpha(z)=1-\sum_{n}ae^{-b(z-n-\frac{1}{2})^{2}}.
\ee
We shall assume that there is a vortex at z=0, and another at
z=2. The superconducting layers are at z=$\pm 1/2$, z=$\pm 3/2$,
z=$\pm 5/2$, etc. We have solved Eqs. (15) and (17) numerically
along the z-axis, in the interval [0,1], assuming solutions
$\psi(\rho)$ and $Q(\rho)$ with $\rho\approx |z|$ and
$\phi\approx\pi/2$.
Since $\Gamma$ is large, our solutions are valid away from the
z-axis as well, in regions with $y\ll\Gamma$. The boundary
conditions are, at the center of the vortex, $\psi(0)=0$,
$Q(0)=-1$, as mentioned earlier, while $\psi(1)=0$, $h(1)=$H at
the edge of the cell. 

\vspace*{3mm}
We repeat here that the boundary conditions at the origin are
actually the boundary conditions arising from Eqs. (13) and (14),
since Eqs. (15) and (17) are valid in [$L/\Gamma$,1] $only$. If
$L$ is much smaller than $\Gamma$ though, we may consider them
as appropriate for Eqs. (15) and (17).

\vspace*{3mm}
We find that for sufficiently high
$d/\lambda$, and for sufficiently large H, the field is inverted
along a substantial interval, but definitely before
$\frac{1}{2}$. This is happening because the slope of $h$ is very
large when $d/\lambda$ and H are large, and hence $h$ drops so
low that it can become negative. We see also that the field is
practically zero along the superconducting
layer, giving thus an almost ideal Meissner effect there. After
crossing the layer, the field $h$ rises again, till it reaches
the value H. In this final interval beyond $\rho=1/2$ we seem to
have the normal state. The behaviour described here can be seen
in Figs. 3, 4, 5, 6, 7. 

\vspace*{3mm}
The curves in Figs. 3 and 4 were obtained for the input
parameters $\Gamma=100$, H=2, $\nu=0.0001$, $a=2.5$, $b=10$,
$d/\lambda=40$, and they correspond to $L=2.2$, $L/\Gamma=0.022$.
The solution holds in [$L/\Gamma$,1] strictly, so $h$ has been
extended towards the origin in Figs. 3a, 5 and 7, using the fact
that it has zero slope there. The curves in Figs. 5 and 6 were
obtained for the input parameters $\Gamma=100$, H=2,
$\nu=0.0001$,
$a=2.5$, $b=10$, $d/\lambda=200$, and they correspond to
$L=2.8$, $L/\Gamma=0.028$. Finally the curve in Fig. 7 has the
input parameters $\Gamma=50$, H=15, $\nu=0.0001$, $a=2.5$,
$b=10$, $d/\lambda=40$, and it corresponds to $L/\Gamma=0.006$.

\vspace*{3mm} 
Note that the field inversion and the
situation described by these Figures, where the vortex at z=0 and
the normal state at z=1 are connected through a superconducting
region, were also discussed recently~\cite{Obu96} in the context
of ordinary superconductors in high fields. It should be noted
furthermore that the above Figures were also obtained by imposing
the boundary conditions $\psi(0)=\psi(1)=0$, $Q(0)=-1$ and
$Q(1)=q$, finding the solution and the corresponding Gibbs free
energy, and then finding which $Q(1)$ gave the minimum Gibbs free
energy. We confirmed thus that the boundary condition $h(1)=$H
arises naturally from the minimization of the Gibbs free energy. 

\vspace*{3mm}
As H goes up, $Q(1)$ and $h(1)$ will
increase. Thus the normal state at the cell boundaries gradually
proceeds inwards, until $Q(1)$ becomes sufficiently large to
destroy the superconductivity on the layers. 

\vspace*{3mm}
It should be mentioned that the above conclusions do not
depend
on the details of
$\alpha(z)$. They depend on the fact that $\alpha(z)$ is even and
periodic, with extrema at 0 and 1, and a layer at $\frac{1}{2}$.
Nor do they involve the y-distance between the vortices, since
$\Gamma$ is large. 

\vspace*{3mm}
We also note that the inversion results from the combination of
two factors: the large negative slope of $h$ at the origin, due
to the large values of $d/\lambda$ and H, and the expulsion of
the field from the layer. It should be therefore quite
independent of the boundary conditions at $z=1$. Indeed the
equation for $h$ has an attractive fixed point at $h=0$.
This is shown clearly in Figs. 3, 5, 7. We see that the field
becomes zero the moment it starts crossing
into the layer, maintaining this value up to the point when it
starts exiting the layer. We have thus a perfect Meissner effect
inside the layer. 
The inversion occurs around the point where the field starts
entering
the layer. It decreases linearly near the origin, with a large
negative slope. This large slope makes it overshoot past zero,
before turning upwards again, towards the zero value of the
Meissner state.  

\vspace*{3mm}
We have checked that the inversion is independent of the boundary
conditions at $z=1$ by solving Eqs. (15) and (17) from $z=0$ to
$z=1/2$, subject to the boundary conditions $Q(0)=-1$,
$\psi(0)=0$, and $\partial h/\partial\rho=0$,
$\partial\psi/\partial\rho=0$ at $\rho=1/2$. We choose
$\Gamma=50$, $H=50$, $\nu=0.0001$, $a=2.5$, $b=10$,
$d/\lambda=40$. The result, plotted in Fig. 8, shows again the
inversion and the attractive fixed point at $h=0$. 

\vspace*{3mm}
We have shown then that in highly anisotropic multilayers, and
for fields parallel to the layers, the field is inverted
away from the center of a vortex, as long as $d\gg\lambda$, i.e.
as long as the vector
potential is able to perform the necessary variations within the
interlayer space, and as long as H is large. This result is
insensitive to the boundary conditions at the edge. 

\vspace*{3mm}
The field inversion discussed here could be seen experimentally
in highly anisotropic multilayers, provided the layers are very
far apart ($d\gg\lambda$), at high fields. Construction of such
multilayers would be especially interesting because it would be
one of the few instances where the magnetic field would have
ample space for varying. In all superconductors examined so far,
even the high T$_{c}$ ones, the penetration depth is much longer
than the interlayer distance.

\vspace*{3mm}
A field inversion has already been presented in [3]. However it
was derived
using a discretized model. The variation of $\Psi$
along the z-axis, which is essential for the inversion to occur,
is introduced in that work by having inequivalent layers, with
various constant order parameter amplitudes. In our work the
variation of $\Psi$ arises naturally from $\alpha(z)$, which
creates the undulating structure. Furthermore, the result of [3]
was
derived near $H_{c1}$, and it does not assume that $d>\lambda$.
Thus the inversion in [3] occurs mostly along the y-axis, unlike
our
inversion.

\vspace*{3mm}
A field inversion has been also presented in [6]. It applies
however to non-layered anisotropic materials, and to tilted
fields. Furthermore, the $\lambda$ there is large, unlike our
$\lambda$.

\vspace*{3mm}
Our field inversion occurs between the vortex and the normal
state. It might occur though, for sufficiently large $d/\lambda$,
even earlier, when the elliptical cells just start touching.
Thus, as the ellipse begins its truncation along the y-axis, it
may be preferable for the field at the edges of the elliptic cell
to become negative. Then the cell could be truncated along the
y-axis without changing its flux! This field inversion would make
possible a drastic reduction of the cell, without affecting the
flux of each cell. Indeed, the flux of the cell would increase
if we were to cut out some of the cell on the right and on the
left, since the field would be negative in those pieces. We would
need then to cut more on the left and on the right, pieces with
positive $h$ now, in order to bring the flux down to its
quantized value. Thus the cell would be truncated considerably
without affecting its flux. If such a field inversion at the
edges of the cell were to occur, then we would have neighboring
vortices with an inverted field in between. In such a case
though, the lattice would be an array
of parallel chains of vortices, and each chain would have a
vortex in every interlayer spacing. The existence of such chains
would require an attraction between the vortices along the
z-direction, an attraction that would be made possible precisely
because of the field inversion.

\newpage
%

\newpage
\noindent
{\bf Figure Captions \hfill}
 
\begin{enumerate}

\item[\bf Figure 1:] 
The vortex lattice at very high fields. Each cell has sides 2$L$
and 2.

\item[\bf Figure 2:] 
A truncated elliptical cell. The distances are always in units
of d.

\item[\bf Figure 3:] 
a) The field $h$ as a function of $\rho$, for H=2, $\Gamma=100$,
$a=2.5$, $b=10$, $d/\lambda=40$, $\nu=0.0001$. b) Detail of this
graph. 

\item[\bf Figure 4:]
The order parameter amplitude $\psi$ as a function of $\rho$, for
the parameters of Fig. 3.

\item[\bf Figure 5:]
The field $h$ as a function of $\rho$, for H=2, $\Gamma=100$,
$a=2.5$, $b=10$, $d/\lambda=200$, $\nu=0.0001$. 

\item[\bf Figure 6:]
The order parameter amplitude $\psi$ as a function of $\rho$,
for the parameters of Fig. 5.

\item[\bf Figure 7:]
The field $h$ as a function of $\rho$, for H=15, $\Gamma=50$,
$a=2.5$, $b=10$, $d/\lambda=40$, $\nu=0.0001$.

\item[\bf Figure 8:]
The field $h$ as a function of $\rho$, for H=50, $\Gamma=50$,
$a=2.5$, $b=10$, $d/\lambda=40$,
$\nu=0.0001$, obtained with the boundary conditions
$\partial\psi/\partial\rho=0$ and $\partial h/\partial\rho=0$ at
$\rho=1/2$.

\end{enumerate}
\end{document}